\documentclass[aps, twocolumn, preprintnumbers, showkeys, tightenlines]{revtex4-1}
\usepackage{graphicx}
\usepackage{amsmath}
\usepackage{amssymb}
\usepackage{amstext}
\usepackage{booktabs}
\usepackage{subfigure}
\usepackage{bm}
\usepackage{color}
\usepackage[singlelinecheck=false]{caption}
\usepackage{natbib}
\usepackage[colorlinks=true,linkcolor=blue,citecolor=blue,urlcolor=blue]{hyperref}

\begin{document}
	
	\preprint{}
	
	\title{Controlled-NOT gate based on the Rydberg states of surface electrons}
	
	\author{Jun Wang}
	\affiliation{Department of Physics, Applied Optics Beijing Area Major Laboratory, Beijing Normal University, Beijing 100875, China}
	
	\author{Wan-Ting He}
	\affiliation{Department of Physics, Applied Optics Beijing Area Major Laboratory, Beijing Normal University, Beijing 100875, China}
	
	\author{Cong-Wei Lu}
	\affiliation{Department of Physics, Applied Optics Beijing Area Major Laboratory, Beijing Normal University, Beijing 100875, China}
	
	\author{Yang-Yang Wang}
	\affiliation{Shaanxi Engineering Research Center of Controllable Neutron Source, School of Electronic Information, Xijing University, Xi'an 710123, China}
	
	\author{Qing Ai}
	\affiliation{Department of Physics, Applied Optics Beijing Area Major Laboratory, Beijing Normal University, Beijing 100875, China}

	\author{Hai-Bo Wang}
	\email{hbwang@bnu.edu.cn}
	\affiliation{Department of Physics, Applied Optics Beijing Area Major Laboratory, Beijing Normal University, Beijing 100875, China}

	\begin{abstract}
		Due to the long coherence time and efficient manipulation, the surface electron (SE) provides a perfect two-dimensional platform for quantum computation and quantum simulation. In this work, a theoretical scheme to realize the controlled-NOT (CNOT) gate is proposed, where the two-qubit system is encoded on the four-level Rydberg structure of SE. The state transfer is achieved by a three-level structure with an intermediate level. By simultaneously driving the SE with two external electromagnetic fields, the dark state in the electromagnetically induced transparency (EIT) effect is exploited to suppress the population of the most dissipative state and increase the robustness against dissipation. The fidelity of the scheme is 0.9989 with experimentally achievable parameters.
	\end{abstract}
	
	
	

	\maketitle
	
	\section{Introduction}\label{sec1}
	
	Universal quantum logic gates \cite{barenco1995pra, sleator1995prl} are the key elements of quantum information processing \cite{chitambar2019rmp} and quantum simulation \cite{buluta2009science, weimer2010np, georgescu2014rmp, mostame2008prl}. In recent years, many schemes of quantum logic gates have been proposed in various physical systems \cite{buluta2011rpp, xiang2013rmp, zhang2017pr, frisk2019nrp}, such as superconducting qubits \cite{makhlin2001rmp, yamamoto2003nature, liu2005prl, you2005pt, you2011nature, gu2017pr}, nuclear magnetic resonance (NMR) systems \cite{jones1998nature, feng2013prl}, cavity quantum electrodynamics (QED) \cite{turchette1995prl, rauschenbeutel1999prl, walther2006rpp, burkard2020nrp}, circuit QED \cite{chiorescu2004nature, dicarlo2009nature, scarlino2019prl}, ion traps \cite{cirac1995prl, poyatos1998prl, figgatt2019nature}, quantum dots \cite{li2003science, ciriano2021prx}, and nitrogen-vacancy centers in diamond \cite{jelezko2004prl, wei2013pra, zhang2020prl}. Among the above proposals, the controlled-NOT (CNOT) gate is one of the most attractive quantum gates, because it can be used to realize universal quantum logic gates with the aid of single-qubit gates \cite{barenco1995pra}. A feasible quantum gate requires the operation time to be shorter than the coherent time of the system, thus the fast manipulation plays the significant role in quantum gates. Some previous works have utilized the electromagnetically induced transparency (EIT) effect \cite{fleischhauer2005rmp, liu2014pra, liu2016pra, gu2016pra, wang2018pra} to reduce the influence of dissipation and accelerate the manipulation \cite{wang2012prl, McDonnell2022prl}. The EIT effect is due to the coherent interference among different transition pathways, which is different from the Autler–Townes splitting \cite{peng2014nc}.

	The surface electron (SE) on the surface of liquid helium provides a controllable two-dimensional (2D) quantum system to study the behavior of strongly-correlated electrons. The SE is attracted by the induced image charge inside the liquid helium and concurrently repulsed by the helium atoms, and therefore the motion perpendicular to the surface is confined and forms a hydrogen-like spectrum \cite{platzman1999science}. Meanwhile, the SE can move freely parallel to the surface, forming a perfect 2D electron system free of the defects and impurities present in semiconductor devices \cite{kawakami2021prl}. The 2D electron system possesses the quantized orbital states when electrons are trapped in an electrostatic potential \cite{koolstra2019natcom}. Both the Rydberg and orbital states can be coupled to the spin states of electrons \cite{kawakami2019prl, schuster2010prl, kawakami2023arxiv}, which have a much longer coherence time than other solid materials \cite{lyon2006pra}, making them an excellent resource for quantum computing. The SE can be manipulated and detected by the circuit QED architecture, which combines the superconducting coplanar-waveguide resonator and the electron trap \cite{koolstra2019natcom, zhou2022nature}. In addition, the SE can also be manipulated and transported through the microchannel devices which are fabricated on the silicon substrate and filled with the superfluid helium \cite{glasson2001prl, ikegami2009prl, rees2011prl, ikegami2012prl, rees2016prl, rees2016prb, badrutdinov2020prl, zou2022njp}. The unprecedented transport efficiency of such microchannel devices \cite{bradbury2011prl} manifests the applications of SE in the large-scale trapped-ion quantum computing \cite{kielpinski2022nature}.

	The highly excited Rydberg state of neutral atoms is a promising candidate for quantum information processing, benefiting from its long coherence time \cite{gallagher1994rydberg} and strong long-range interactions \cite{saffman2010rmp}. The SE system can be used to simulate Rydberg states \cite{platzman1999science} because it has the same hydrogen-like energy spectrum as Rydberg atoms. At low temperatures, the dissipation of SE is mainly due to the height variations of the helium surface, which can be quantized as ripplons \cite{kawakami2021prl}. The lifetime $T_1$ exceeds 10~$\mu$s at 10~mK, which is sufficiently long compared to the Rabi frequency $\Omega$, i.e., $\Omega T_1>10^4$ \cite{platzman1999science}. A more realistic lifetime measured by some recent works \cite{monarkha2006ltp, monarkha2007jltp, monarkha2010ltp, kawakami2021prl} is 1~$\mu$s, which is adopted by our work.

	Here, we present a scheme to realize the CNOT gate in a single SE system. We encode the two-qubit system in the four-level Rydberg structure of SE. The first qubit represents whether the electron is close or far away from the liquid surface, i.e., whether the electron is in the ``lower'' or ``upper'' mode. Each mode comprises a two-level system, which is labeled by the second qubit. This proposal is analogous to the hyperentanglement that combines several degrees of freedom (DOFs) of a single particle, such as the spatial mode and the polarization of a single photon \cite{barreiro2005prl, yang2005prl, deng2017sb}. Although the multilevel structure is difficult to be scaled up on a single electron, it can be scaled up with the assistance of adjacent electrons or different DOFs. The dipole-dipole interaction of neutral atoms has been used to realize the quantum gates \cite{Jaksch2000prl, McDonnell2022prl}. Recent work based on the SE has proposed a scheme to couple two adjacent electrons and scale up the system via the dipole-dipole interaction of electron in different Rydberg states \cite{kawakami2023arxiv}. Meanwhile, the Rydberg states and the spin states of the SE can be coupled via an inhomogeneous magnetic field \cite{tokura2006prl, kawakami2023arxiv}.	On the other hand, the SE is also a promising platform for quantum simulation \cite{buluta2009science, weimer2010np, georgescu2014rmp, mostame2008prl}, and the four-level structure in our scheme can be used to simulate the quantum coherent effects of multilevel molecules such as the four-level pigment–protein molecules in photosynthetic light harvesting \cite{wang2018NPJQI}. 	
	
	Since the highly excited Rydberg states are sensitive to the frequency fluctuation of the driving fields \cite{saffman2010rmp} and their level spacing is narrow, the direct driving for state transfer could easily cause the undesirable transitions to other neighboring states. Therefore, to accurately achieve the transition between $|10\rangle$ and $|11\rangle$, we use an intermediate level to avoid the undesired transitions to other highly excited states. By applying two driving pulses simultaneously, we exploit the dark state in the EIT effect \cite{fleischhauer2005rmp,wangyy2018pra} to reduce the population of the most dissipative intermediate level and increase the robustness against dissipation.

	This paper is organized as follows. In Sec.~\ref{sec2}, we describe the coherent-driving scheme based on the Rydberg states of SE. In Sec.~\ref{sec3}, we compare our scheme with other schemes and show how the CNOT gate was accelerated by the two simultaneous driving fields at the same time. We also investigate the effects of detuning and dissipation on the fidelity. Finally, we conclude the work and give a prospect in Sec.~\ref{sec4}. In Appendix~\ref{sec:AppA}, we analyze the decay mechanism of the excited states. In Appendix~\ref{sec:AppB}, we provide the eigenvalues of the non-Hermitian Hamiltonian by the perturbation theory.

	\section{The model} \label{sec2}
	\begin{figure}[!ht]
		\centering
		\includegraphics[width=1\linewidth]{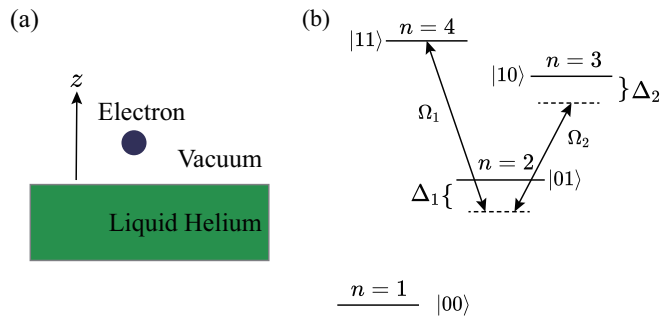}
		\caption{Schematic diagram of the CNOT gate based on SE. (a) The SE on the surface of liquid helium. (b) The two-qubit system is encoded in the four-level SE Rydberg states, where two coherent driving fields with frequencies $\omega_j$ and Rabi frequencies $\Omega_j$ ($j=1,2$) are applied simultaneously. Here, $\Delta_1$ and $\Delta_2$ are the single-photon and two-photon detunings,  respectively. }\label{fig1}
	\end{figure}
	
	As shown in Figure~\ref{fig1}(a), along the direction $z$ perpendicular to the interface, the SE is attracted by the induced image charge inside the liquid helium and concurrently repulsed by the helium atoms, and therefore the motion of the SE is confined by the hydrogen-like potential $V=-\Lambda e^2/z$ for $z>0$, where $e$ is the charge of the electron and $\Lambda=(\epsilon-1)/[4(\epsilon+1)]$ with the dielectric constant $\epsilon\approx 1.057$. The quantized SE states possess the hydrogen-like energy spectrum as \cite{platzman1999science, monarkha2004two}
	\begin{align}
		\varepsilon_n^{(\perp)}=-\frac{m_e e^4\Lambda^2}{2\hbar^2 n^2}=-\frac{R}{n^2}, \label{eq1}
	\end{align}
	where $m_e$ is the mass of the electron, the positive integer $n$ labels the SE state, and $R\approx0.7$~meV \cite{platzman1999science} is the Rydberg energy. The wave function of SE is \cite{nieto2000pra}	
	\begin{align}
		\psi_n(z)=2n^{-5/2}r_B^{-3/2}z\exp\left(-\frac{z}{nr_B}\right)L_{n-1}^{(1)}\left(\frac{2z}{nr_B}\right),\label{eq2}
	\end{align}
	where $r_B=\hbar^2/(m_e e^2\Lambda)$ is the effective Bohr radius, and
	\begin{align}
		L_{n}^{(\alpha)}(x)=\frac{e^x x^{-\alpha}}{n!}\frac{d^n}{dx^n}(e^{-x}x^{n+\alpha})
	\end{align}
	is the Laguerre polynomial.
	
	The expected positions $\langle z \rangle_n=\langle \psi_n|z|\psi_n \rangle$ of the electrons in different Rydberg states $|\psi_n \rangle$ are different. The electrons in higher excited states are further away from the liquid surface. For example, $\langle z \rangle_2/\langle z \rangle_1=4$, $\langle z \rangle_3/\langle z \rangle_1=9$, $\langle z \rangle_4/\langle z \rangle_1=16$ and so on. This phenomenon is remarkable for highly excited Rydberg states. The expected position $\langle z \rangle$ has practical applications in physical systems. For example, the electron with larger $\langle z \rangle$ has larger electric dipole moment and induces larger dipole-dipole interaction with the adjacent electrons, and thus can be used to couple adjacent electrons \cite{Jaksch2000prl, McDonnell2022prl}. Meanwhile, in the gradient magnetic field, the electrons with different $\langle z \rangle$ have different spin-level shifts due to the Zeeman effect, and thus can be used to couple the Rydberg states with the spin states \cite{tokura2006prl, kawakami2023arxiv}.
	
	Encoding a two-qubit system on a single electron is analogous to the hyperentanglement which combines several DOFs of a single particle \cite{barreiro2005prl, yang2005prl, deng2017sb}. We consider the ground state and the first excited state of the SE as the ``lower'' mode and consider the highly excited Rydberg states as the ``upper'' mode. The first bit indicates which mode the SE is in. The ``lower'' mode is labeled as $|0\rangle$, and the  ``upper'' mode is labeled as $|1\rangle$. Each mode comprises a two-level system, which is labeled by the second bit. Therefore, the ground state with $n=1$ is labeled as $|00\rangle$, the first excited state with $n=2$ is labeled as $|01\rangle$, the second excited state with $n=3$  is labeled as $|10\rangle$, and the third excited state with $n=4$ is labeled as $|11\rangle$, as shown in Figure~\ref{fig1}(b). The CNOT gate in our scheme means that the second bit flips only if the state is in the ``upper'' mode, while the second bit doesn't flip if the state is in the ``lower'' mode.
	
	The level spacing of highly excited Rydberg states is narrow because the energy space $\Delta E_n$ decreases with $n$. Moreover, since the dipole moment matrix elements for the transitions between highly excited states are large, the direct transition between these states is sensitive to the frequency fluctuation of the driving fields \cite{saffman2010rmp}. This sensitivity of the direct manipulation of highly excited states could easily cause the undesirable transition to other neighboring states. Therefore, we use the first excited state $|01\rangle$ as an intermediate state to realize the state swap between $|10\rangle$ and $|11\rangle$. 	
	
	Because the first excited state $|01\rangle$ is the most dissipative state (see Appendix.~\ref{sec:AppA}), we exploit the dark state in the EIT effect to reduce the population on $|01\rangle$. Two driving fields with frequencies $\omega_1$ ($\omega_2$) and Rabi frequencies $\Omega_1$ ($\Omega_2$) are used to drive the transition $|01\rangle\rightleftharpoons|11\rangle$ ($|01\rangle\rightleftharpoons|10\rangle$), as shown in Figure~\ref{fig1}(b). The single-photon detuning is $\Delta_1=\omega_1-(\omega_{11}-\omega_{01})$ and the two-photon detuning is $\Delta_2=\omega_1-\omega_2-(\omega_{11}-\omega_{10})$, where $\omega_{01}$, $\omega_{10}$ and $\omega_{11}$ are the energies of the states $|01\rangle$, $|10\rangle$ and $|11\rangle$ respectively. As the driving frequencies are far detuned from the transitions between $|00\rangle$ and other states, we consider $|00\rangle$ to be a decoupled state. In the subspace spanned by $\{|01\rangle, |10\rangle, |11\rangle\}$, the Hamiltonian is
	\begin{align}
		H=~&\omega_{01}|01\rangle\langle01|+\omega_{10}|10\rangle\langle 10|+\omega_{11}|11\rangle\langle11| \nonumber\\
		&-\Omega_1\cos\omega_1t(|01\rangle\langle11|+|11\rangle\langle 01|) \nonumber\\
		&-\Omega_2\cos\omega_2t(|01\rangle\langle10|+|10\rangle\langle 01|),	
	\end{align}
	where $\hbar=1$. In the rotating frame with driving frequencies $U=\exp[i\omega_1t|01\rangle\langle01|+i(\omega_1-\omega_2)t|10\rangle\langle 10|]$, under the rotating-wave approximation \cite{Scully1997quantum, Ai2010PRA}, and taking $\omega_{11}=0$ as the zero point of energy, the matrix form of the Hamiltonian reads
	\begin{align}
		H_I=~&i\frac{dU^{\dag}}{dt}U+U^{\dag}HU \nonumber \\
		=~&-\frac{1}{2}\begin{pmatrix}
			-2\Delta_1 & \Omega_2 & \Omega_1 \\
			\Omega_2 & -2\Delta_2 & 0 \\
			\Omega_1 & 0 & 0
		\end{pmatrix}.
	\end{align}
	
	The evolution of the system can be described by the quantum master equation \cite{Breuer2002}
	\begin{align}
		\frac{\partial}{\partial t}\rho=-i[H_I,\rho]-\mathcal{L}(\rho),
	\end{align}
	where the Lindblad operator is
	\begin{align}
		\mathcal{L}(\rho)=~&\kappa_1[|01\rangle\langle01| \rho |01\rangle\langle01| -\frac{1}{2}\{|01\rangle\langle01|,\rho\}] \nonumber\\
		&+ \kappa_2[|10\rangle\langle10|\rho|10\rangle\langle10| -\frac{1}{2}\{|10\rangle\langle10|,\rho\}] \nonumber\\
		&+ \kappa_3[|11\rangle\langle11|\rho|11\rangle\langle11| -\frac{1}{2}\{|11\rangle\langle11|,\rho\}] \label{eq7}
	\end{align}
	with $\{A,B\}=AB+BA$ being the anti-commutator, and $\kappa_1$, $\kappa_2$ and $\kappa_3$ being the decay rates of the states $|01\rangle$, $|10\rangle$ and $|11\rangle$ respectively. In our scheme, the electron is only confined by the image potential and no electric holding field is applied. In this case, the decay rates of the Rydberg states of the SE decrease with $n$, cf. Appendix~\ref{sec:AppA}. Thus, in the analytic calculation, we mainly consider the dissipations of the energy levels $|01\rangle$ and $|10\rangle$ and neglect the dissipation of $|11\rangle$, while all of the dissipations are considered in the numerical simulation. It is noteworthy that in experiments there is usually an electric holding field $E_z$ applied perpendicular to the liquid surface. For the experimental configuration such as $E_z\approx200$ V/cm \cite{kawakami2023arxiv} and $E_z\approx1$ kV/cm \cite{bradbury2011prl, zou2022njp}, the decay rates of the Rydberg states of the SE increase with $n$, cf. Appendix~\ref{sec:AppA}. Recent works have provided efficient quantum algorithms to simulate the quantum open system both theoretically \cite{zhang2021FoP} and experimentally \cite{wang2018NPJQI}, even for the non-Markovian process \cite{chen2022NPJQI}.
	
	To solve the time evolution analytically, we neglect the quantum jump term and describe the evolution by the Schr\"{o}dinger equation with the following non-Hermitian Hamiltonian
	\begin{align}
		H_I^{d}=~&H_I-i\frac{\kappa_1}{2}|01\rangle\langle 01| -i\frac{\kappa_2}{2}|10\rangle\langle 10| \nonumber\\
		=~&-\frac{1}{2}\begin{pmatrix}
			-2\delta_1 & \Omega_2 & \Omega_1 \\
			\Omega_2 & -2\delta_2 & 0 \\
			\Omega_1 & 0 & 0
		\end{pmatrix}, \label{eq6}
	\end{align}
	where $\delta_1=\Delta_1-i\kappa_1/2$, $\delta_2=\Delta_2-i\kappa_2/2$. When $\delta_2\ll\Omega_1, \Omega_2$, we take $\delta_2$ as the perturbation term. The approximated eigenstates by taking $\delta_2=0$ are \cite{fleischhauer2005rmp}
	\begin{align}
		\begin{split}
			&|a_1\rangle=-\sin\theta|10\rangle+\cos\theta|11\rangle, \\
			&|a_2\rangle=\cos\phi|01\rangle+\cos\theta\sin\phi|10\rangle+\sin\theta\sin\phi|11\rangle, \\
			&|a_3\rangle=-\sin\phi|01\rangle+\cos\theta\cos\phi|10\rangle+\sin\theta\cos\phi|11\rangle,
		\end{split}
	\end{align}
	where the mixing angles are $\tan\theta=\Omega_1/\Omega_2$ and $\tan 2\phi=\Omega/\delta_1$, and $\Omega=\sqrt{\Omega_1^2+\Omega_2^2}$. It is noteworthy that $|a_1\rangle$ is a dark state since there is no population on the state $|01\rangle$. As shown in Appendix~\ref{sec:AppB}, the first-order approximations of the eigenvalues are
	\begin{align}
		\begin{split}
			E_1\simeq&\frac{\Omega_1^2}{\Omega^2}\delta_2, \\
			E_2\simeq&\frac{\Omega}{2}+\frac{\delta_1}{2}+\frac{\Omega_2^2}{2\Omega^2}\delta_2, \\
			E_3\simeq&-\frac{\Omega}{2}+\frac{\delta_1}{2}+\frac{\Omega_2^2}{2\Omega^2}\delta_2,
		\end{split}
	\end{align}
	The time evolution of the initial state $|\psi(0)\rangle=C_1|a_1\rangle+C_2|a_2\rangle+C_3|a_3\rangle$ is
	\begin{align}
		|\psi(t)\rangle=~&C_1e^{-iE_1t}|a_1\rangle+C_2e^{-iE_2t}|a_2\rangle+C_3e^{-iE_3t}|a_3\rangle \nonumber \\
		=~&(C_2\cos\phi e^{-iE_2t} -C_3\sin\phi e^{-iE_3t})|01\rangle \nonumber \\
		&+(-C_1\sin\theta e^{-iE_1t} +C_2\cos\theta\sin\phi e^{-iE_2t} \nonumber \\
		&+C_3\cos\theta\cos\phi e^{-iE_3t})|10\rangle \nonumber \\
		&+(C_1\cos\theta e^{-iE_1t} +C_2\sin\theta\sin\phi e^{-iE_2t}  \nonumber\\
		&+C_3\sin\theta\cos\phi e^{-iE_3t})|11\rangle.
	\end{align}
	
	The key point of the CNOT gate scheme is to swap the population of $|10\rangle$ and $|11\rangle$, while maintaining the low population of $|01\rangle$ in the final state. The state swapping is based on the population oscillation. As $E_1\ll E_2,~E_3$, the main oscillation factors are $\exp(-iE_2t)$ and $\exp(-iE_3t)$. To achieve the maximum population reversal of $|10\rangle$ and $|11\rangle$, these two oscillation terms $\exp(-iE_2t)$ and $\exp(-iE_3t)$ should have the same period. The synchronization of $\exp(-iE_2t)$ and $\exp(-iE_3t)$ requires $E_2=-E_3$, so $\delta_1$ and $\delta_2$ must be much smaller than $\Omega$, i.e.,
	\begin{align}
		\tan\phi=1. \label{eq10}
	\end{align}
	Meanwhile, the oscillation term in the coefficient of $|10\rangle$ and $|11\rangle$ must be the same, which means that $\cos\theta=\sin\theta$, i.e.,
	\begin{align}
		\Omega_1=\Omega_2.
	\end{align}
	
	\begin{figure}[!ht]
		\centering
		\includegraphics[width=1\linewidth]{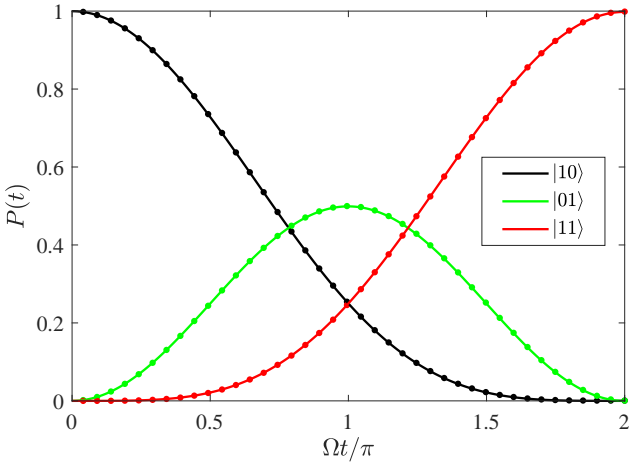}
		\caption{The populations $P(t)$ of different states with the initial state $|10\rangle$ and the dissipation rates $\kappa_1/\Omega=10^{-3}$,  $\kappa_2=0.3439\kappa_1$ and $\kappa_3=0.1520\kappa_1$. The solid lines are the analytic solutions and the dots are the numerical solutions by the quantum master equation Eq.~(\ref{eq7}). }\label{fig2}
	\end{figure}
	
	Under these conditions, if the initial state is $|\psi(0)\rangle=|10\rangle$, the final state is
	\begin{align}
		|\psi(t)\rangle=&
		\left(\frac{1}{2}\cos\frac{\Omega t}{2} e^{-i\frac{2\delta_1+\delta_2}{4}t}+\frac{1}{2}e^{-i\frac{\delta_2}{2}t}\right)|10\rangle \nonumber\\
		&-\frac{i}{\sqrt{2}}\sin\frac{\Omega t}{2}e^{-i\frac{2\delta_1+\delta_2}{4}t}|01\rangle \nonumber\\
		&+\left(\frac{1}{2}\cos\frac{\Omega t}{2} e^{-i\frac{2\delta_1+\delta_2}{4}t}-\frac{1}{2}e^{-i\frac{\delta_2}{2}t}\right)|11\rangle. \label{eq12}
	\end{align}
	If the initial state $|\psi(0)\rangle=|01\rangle$, the final state is
	\begin{align}
		|\psi(t)\rangle=&
		-\frac{i}{\sqrt{2}}\sin\frac{\Omega t}{2}e^{-i\frac{2\delta_1+\delta_2}{4}t}|10\rangle \nonumber\\
		&+\cos\frac{\Omega t}{2} e^{-i\frac{2\delta_1+\delta_2}{4}t}|01\rangle \nonumber\\
		&-\frac{i}{\sqrt{2}}\sin\frac{\Omega t}{2}e^{-i\frac{2\delta_1+\delta_2}{4}t}|11\rangle. \label{eq13}
	\end{align}
	If the initial state $|\psi(0)\rangle=|11\rangle$, the final state is
	\begin{align}
		|\psi(t)\rangle=&
		\left(\frac{1}{2}\cos\frac{\Omega t}{2} e^{-i\frac{2\delta_1+\delta_2}{4}t}-\frac{1}{2}e^{-i\frac{\delta_2}{2}t}\right)|10\rangle \nonumber\\
		&-\frac{i}{\sqrt{2}}\sin\frac{\Omega t}{2}e^{-i\frac{2\delta_1+\delta_2}{4}t}|01\rangle \nonumber\\
		&+\left(\frac{1}{2}\cos\frac{\Omega t}{2} e^{-i\frac{2\delta_1+\delta_2}{4}t}+\frac{1}{2}e^{-i\frac{\delta_2}{2}t}\right)|11\rangle. \label{eq14}
	\end{align}
	
	The time evolution of different initial states is shown in Figure~\ref{fig2}. The Rabi frequency $\Omega=10^9~\textrm{s}^{-1}$ \cite{platzman1999science} and the decay rate $\kappa_1=10^6~\textrm{s}^{-1}$ \cite{monarkha2007jltp, kawakami2021prl}, thus $\kappa_1/\Omega=10^{-3}$. We take $\kappa_2=0.3439\kappa_1$ and $\kappa_3=0.1520\kappa_1$, according to Appendix~\ref{sec:AppA}. The maximum population reversal between $|10\rangle$ and $|11\rangle$ is reached when $\Omega t=2\pi$. The analytical solutions are in good agreement with the numerical solutions calculated by Qutip \cite{Johansson2012CPC,Johansson2013CPC}.

	\section{Fidelity analysis} \label{sec3}
	
	\begin{figure}[!ht]
		\centering
		\includegraphics[width=1\linewidth]{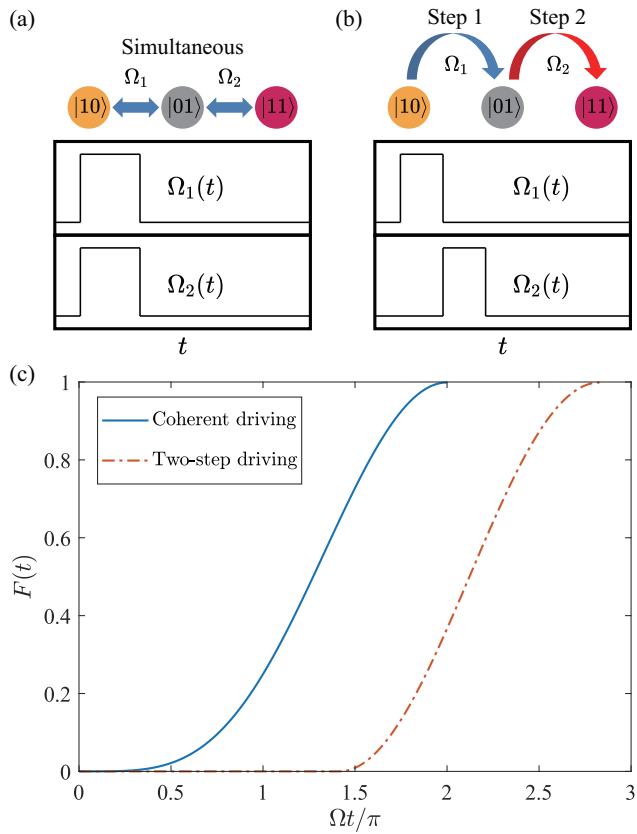}
		\caption{(a) Schematic diagram of the coherent-driving scheme. (b) Schematic diagram of the two-step driving scheme. (c) The fidelity $F$ of the state transfer $|10\rangle\to|11\rangle$. The dissipation rates are the same as Figure~\ref{fig2}.  }\label{fig3}
	\end{figure}
	
	Since the driving frequencies are far detuned from the transition between $|00\rangle$ and other states, we consider $|00\rangle$ to be a decoupled state. From the time evolution in Eq.~(\ref{eq12})-(\ref{eq14}), the maximum population reversal between $|10\rangle$ and $|11\rangle$ is achieved when $\Omega t=2\pi$. The state fidelity $F$ between the final state $\rho(t)$ and the ideal target state $\rho_i$ is defined as \cite{uhlmann1976rmp, Jozsa1994jmo}
	\begin{align}
		F=\left(\textrm{Tr}\sqrt{\sqrt{\rho_i}\rho(t)\sqrt{\rho_i}}\right)^2.
	\end{align}
	In our coherent-driving scheme, two driving fields interact with the SE simultaneously, with the dark and bright states being used equally for the state transfer. The maximum fidelity is achieved when $t=2\pi/\Omega$, as shown in Figure~\ref{fig3}(c). For comparison, we show the result of the two-step driving scheme in Figure~\ref{fig3}(b). We derive the population reversal between $|10\rangle$ and $|01\rangle$ by the first driving pulse $\Omega_1(t)$, and then derive the population reversal between $|01\rangle$ and $|11\rangle$ by the second driving pulse $\Omega_2(t)$. The maximum fidelity is achieved when $t=2\sqrt{2}\pi/\Omega$, which is longer than the coherent-driving scheme. In addition, the two-step driving scheme can only achieve the one-way state transfer based on the driving pulse sequence, but a NOT-gate requires the bidirectional transfer with the same driving pulse sequence.

\begin{figure}[!h]
		\centering
		\includegraphics[width=1\linewidth]{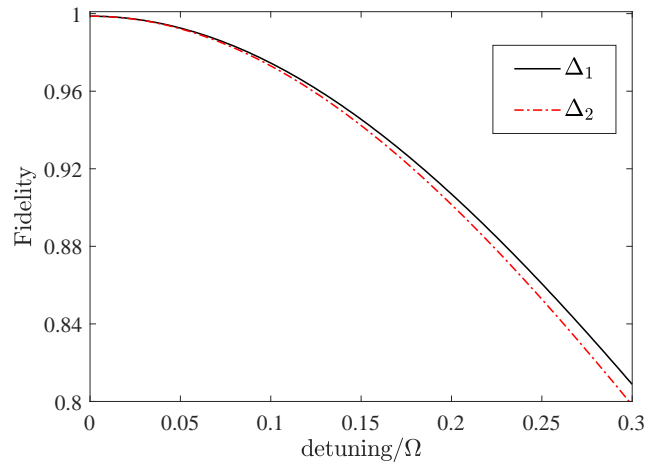}
		\caption{State fidelity versus the single-photon detuning $\Delta_1$ and the two-photon detuning $\Delta_2$. The dissipation rates are the same as Figure~\ref{fig2}. The solid (solid-dot) line shows the dependence on $\Delta_1$ ($\Delta_2$) with $\Delta_2=0$ ($\Delta_1=0$). }\label{fig4}
	\end{figure}

	In Sec.~\ref{sec2}, we have found that $\Delta_1$ need to be much smaller than $\Omega$ in order to synchronize the oscillation terms $\exp(-iE_2t)$ and $\exp(-iE_3t)$. When the single-photon resonance condition is invalid, the fidelity of the state transfer decreases, as shown in Figure \ref{fig4}. It is noteworthy that the oscillation period changes with $\Delta_1$. Thus, the fidelity is calculated at the maximum in the first period. On the other hand, when $\Delta_2$ becomes large, the perturbation method is invalid, but the evolution  of the states can still be derived from the master equation. As shown in Figure~\ref{fig4}, when the two-photon resonance condition is invalid, the fidelity of the state transfer decreases because the EIT effect is suppressed.
	
\begin{figure}[!h]
		\centering
		\includegraphics[width=1\linewidth]{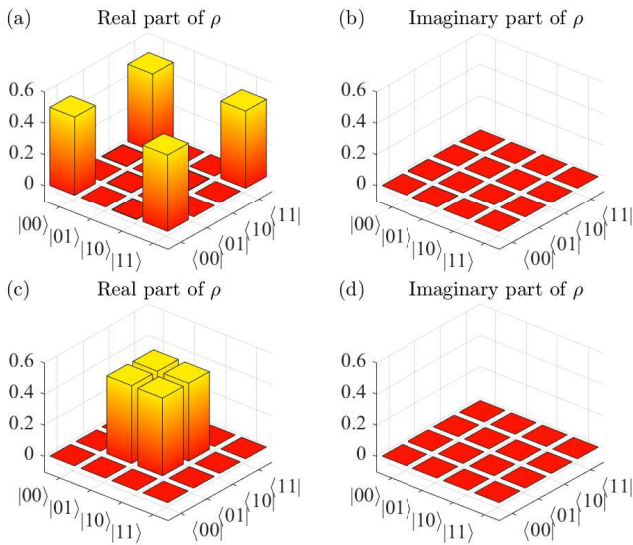}
		\caption{The matrix elements of the output density operator when the initial state is $(|0\rangle+|1\rangle)\otimes|0\rangle/\sqrt{2}$ for (a) and (b), and the initial state is $(|0\rangle+|1\rangle)\otimes|1\rangle/\sqrt{2}$ for (c) and (d). The dissipation rates are the same as Figure~\ref{fig2}. }\label{fig5}
	\end{figure}

	Figure~\ref{fig3}(c) only shows the fidelity of the state transfer with the input state $|10\rangle$. In Table \ref{tab1} we present the output state fidelities of the CNOT gate under typical input states. When the control bit is $|0\rangle$, the target bit remains in the initial state. When the control bit is $|1\rangle$, the target bit flips. Meanwhile, when the input state is a superposition state, the output also corresponds to the characteristic of the CNOT gate. The output density matrices of $(|0\rangle+|1\rangle)\otimes|0\rangle/\sqrt{2}$ and  $(|0\rangle+|1\rangle)\otimes|1\rangle/\sqrt{2}$ are shown in Figure~\ref{fig5}.

	\begin{table}[!h]
		\centering
		\renewcommand\arraystretch{1.5}
		\tabcolsep=0.3cm
		\caption{The output state fidelities of the CNOT gate under typical input states. The dissipation rates are the same as Figure~\ref{fig2}. }
		\begin{tabular}{c c l}
			\toprule[1.5pt]
			Input state & Ideal output state & Fidelity \\
			\midrule[1pt]
			$|00\rangle$ & $|00\rangle$ & 1 \\
			$|01\rangle$ & $|01\rangle$ & 0.9987 \\
			$|10\rangle$ & $|11\rangle$ & 0.9987 \\
			$|11\rangle$ & $|10\rangle$ & 0.9987 \\
			$(|0\rangle+|1\rangle)\otimes|0\rangle/\sqrt{2}$ & $(|00\rangle+|11\rangle)/\sqrt{2}$ & 0.9990 \\
			$(|0\rangle+|1\rangle)\otimes|1\rangle/\sqrt{2}$ & $(|01\rangle+|10\rangle)/\sqrt{2}$ & 0.9980 \\
			\bottomrule[1.5pt]
		\end{tabular}\label{tab1}
	\end{table}

	To demonstrate the characteristic of the entire gate, we calculate the gate fidelity which is defined as \cite{Palao2002prl, Wu2017ieee}
	\begin{align}
		F=\frac{1}{N}|\textrm{Tr}(e^{i\phi}U_{r}^{\dag})U_{i}|,
	\end{align}
	where $N$ is the dimension of the Hilbert space, $U_{i}$ is the ideal gate operation, $U_{r}$ is the real operation in our scheme, and $\phi$ is a global phase to maximize $F$. Since the transitions between $|00\rangle$ and other states are negligible, we can analytically derive the operation matrix of the CNOT gate in our scheme from the evolution of different initial states:
	\begin{align}
		U_{r}=&\begin{pmatrix}
			1 & 0 & 0 &0  \\
			0 & a & 0 &0  \\
			0 & 0 & b &c  \\
			0 & 0 & c &b
		\end{pmatrix},
	\end{align}
	where
	\begin{align}
		\begin{split}	&a=e^{-\frac{(2\kappa_1+\kappa_2)\pi}{4\Omega}}, \\ &b=\frac{1}{2}e^{-\frac{(2\kappa_1+\kappa_2)\pi}{4\Omega}}-\frac{1}{2}e^{-\frac{\kappa_2\pi}{2\Omega}}, \\ &c=\frac{1}{2}e^{-\frac{(2\kappa_1+\kappa_2)\pi}{4\Omega}}+\frac{1}{2}e^{-\frac{\kappa_2\pi}{2\Omega}}.
		\end{split}
	\end{align}
	This formula is obtained by an additional phase operation on $|01\rangle$, $|10\rangle$ and $|11\rangle$, which adds a $\pi$ phase to these three energy levels. Compared to the ideal CNOT gate
	\begin{align}
		U_{i}=&\begin{pmatrix}
			1 & 0 & 0 & 0  \\
			0 & 1 & 0 & 0  \\
			0 & 0 & 0 & 1  \\
			0 & 0 & 1 & 0
		\end{pmatrix},
	\end{align}
	the fidelity of $U_{r}$ is
	\begin{align}
		F=\frac{1}{4}|\textrm{Tr}(U_r^T U_i)|=\frac{1+a}{4}+\frac{c}{2}.
	\end{align}
	
\begin{figure}[!h]
		\centering
		\includegraphics[width=1\linewidth]{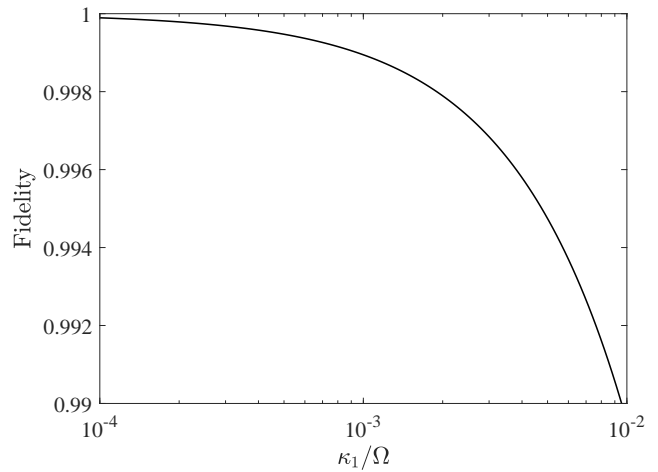}
		\caption{The gate fidelity between the ideal CNOT gate and the gate in our scheme under different dissipation rates. }\label{fig6}
	\end{figure}

	Experimental systems usually apply a vertical static electric field $\vec{E}=E_{\perp}\vec{e}_z$ in order to tune the energy spacing between Rydberg states. The lifetime of the excited state decreases with $E_{\perp}$, and for typical $E_{\perp}=100$ V/cm it becomes 5 times shorter \cite{monarkha2007jltp}. Thus, in Figure~\ref{fig6} we analyze the gate fidelity under different dissipation rates. The fidelity $F>0.99$ in a wide range of $\kappa_1$, and $F=0.9989$ with parameters $\kappa_1/\Omega=10^{-3}$ which is experimentally achievable \cite{platzman1999science,  monarkha2007jltp, kawakami2021prl}.
	
	\section{Conclusion and remarks}\label{sec4}
	
	In this work, we present a scheme to realize the CNOT gate in the four-level Rydberg structure of SE. We use a three-level structure to realize the state transfer. By applying two driving pulses simultaneously, we exploit the dark state in the EIT effect to suppress the population of the most-dissipative state and increase the robustness against dissipation \cite{Huang2022AdP}. We obtain the time evolution of the system both analytically and numerically. We optimize the Rabi frequencies and the detunings of the driving fields to achieve the maximum population exchange of $|10\rangle$ and $|11\rangle$. The optimal state transfer requires that both the single-photon and the two-photon resonances are satisfied. We also calculate the fidelity of the state transfer and of the entire gate. The fidelity is 0.9989 with experimentally achievable parameters.
	
	The CNOT operation involves a three-level structure whose configuration can be the ladder type, $V$ type, or $\Lambda$-type \cite{fleischhauer2005rmp}. The choice of configuration depends on the relative dissipation rates of the three levels. In our scheme, we mainly consider the case without the electric holding field. In this case the decay rates of the Rydberg states decrease with $n$. Therefore, we choose the $V$-type configuration to consider the most dissipative state as the intermediate state and use the EIT effect to suppress the population on this state. For the case with a considerable electric holding field, the decay rates are significantly different from those in the zero-field case, and the $\Lambda$-type configuration of the three-level structure should be chosen accordingly. 
	
	It's noteworthy that both the iSWAP gate and CNOT gate can be used to construct universal quantum computation. The efficiencies of these two gates are determined by the character of the hardware. For example, in spin systems, it's easier to realize the XY interaction by spin-spin interaction. Thus, the iSWAP gate is more efficient in this circumstance \cite{Lidar2001prl, Schuch2003pra, Tanamoto2009prl}. In our scheme we only use the dipole transition of the surface electron, and thus iSWAP gates do not have the advantage of CNOT gates. If the spin-spin interaction of surface electrons is used for quantum computation, the iSWAP gate may be a more efficient candidate.
	
	In our scheme, the first logical bit of the two-bit system labels whether the electron is close or far away from the surface because the expected positions of different Rydberg states are different. The expected position $\langle z \rangle$ has practical applications in physical systems. Recent works have used the expected position of electrons to scale up the system. For example, the electron with larger $\langle z \rangle$ has larger electric dipole moment and induces larger dipole-dipole interaction with the adjacent electrons, and thus can be used to couple adjacent electrons \cite{Jaksch2000prl, McDonnell2022prl}. Meanwhile, in the gradient magnetic field, the electrons with different $\langle z \rangle$ have different spin-level shifts due to the Zeeman effect, and thus can be used to couple the Rydberg states with the spin states \cite{tokura2006prl, kawakami2023arxiv}. Our scheme demonstrates the advantages of precise manipulation on highly excited Rydberg states with narrow energy space and therefore provides potential applications in the schemes based on the dipole-dipole interaction of highly excited Rydberg states and the coupling between the Rydberg states and the spin states.

	\begin{acknowledgements}
		This work is supported by the National Natural Science Foundation of China under Grant No.~61675028 and the Interdiscipline Research Funds of Beijing Normal University. Q. Ai is supported by the Beijing Natural Science Foundation under Grant No.~1202017 and the National Natural Science Foundation of China under Grant Nos.~11674033, 11505007, and Beijing Normal University under Grant No.~2022129. Y. Y. Wang is supported by the Natural Science Basic Research Program of Shaanxi under Grant No.~2023-JC-QN-0092.
	\end{acknowledgements}

	\appendix	
	
	\section{Decay mechanism of the surface electron}
	\label{sec:AppA}
	The two-ripplon scattering in the short-wavelength range dominates the decay mechanism at low temperature, and the decay rate of the first excited state ($n=2$) is expressed as \cite{monarkha2006ltp, monarkha2007jltp, monarkha2010ltp}
	\begin{align}
		\frac{1}{\tau_{2,k}}=\frac{m_e\kappa_0^2}{4\pi\hbar\alpha\rho}\left(\frac{\rho}{4\hbar^2\alpha}\right)^{1/3}\left(\frac{\partial V}{\partial z}\right)_{11}\left(\frac{\partial V}{\partial z}\right)_{22}\Delta_{21}^{2/3},
	\end{align}
	where $V(z)$ is the electron potential, $\kappa_0$ is the penetration depth of the electron wave function into liquid, $\alpha$ is the surface tension of liquid helium, and $\rho$ is the liquid mass density. The decay rate is determined by the diagonal matrix element of $\partial V/\partial z$ and the energy difference $\Delta$ between the initial state and the lower-lying Rydberg states that the initial state leaks to.
	
	Since the spontaneous two-ripplon emission process decreases the energy of surface electrons, we neglect the leakage to the higher-lying Rydberg states when calculating the decay rate. The decay rate of the highly-excited Rydberg state is the sum of the leakage to all lower-lying Rydberg states, i.e., the decay rate $\kappa^{(n)}$ of the Rydberg state with the quantum number $n$ is
	\begin{align}
		\kappa^{(n)}=K\sum_{l=1}^{n-1}\left(\frac{\partial V}{\partial z}\right)_{ll}\left(\frac{\partial V}{\partial z}\right)_{nn}\Delta_{nl}^{2/3}, \label{eqA2}
	\end{align}
	where $K=\frac{m_e\kappa_0^2}{4\pi\hbar\alpha\rho}(\frac{\rho}{4\hbar^2\alpha})^{1/3}$ is a constant.
	
	\begin{table}[!h]
		\centering
		\renewcommand\arraystretch{1.5}
		\tabcolsep=0.2cm
		\caption{The decay rates of five lowest excited states under the electric holding field. Note that the dependence of $\kappa^{(n)}$ on $n$ is significantly different for the case with and without the applied electric field.}
		\begin{tabular}{c c c c c}
			\toprule[1.5pt]
			$E_z$ (V/cm) & $\kappa^{(3)}/\kappa^{(2)}$ & $\kappa^{(4)}/\kappa^{(2)}$ & $\kappa^{(5)}/\kappa^{(2)}$ & $\kappa^{(6)}/\kappa^{(2)}$ \\
			\midrule[1pt]
			0 & 0.3439 & 0.1520 & 0.0795 & 0.0465 \\
			100 & 0.9807 & 1.0050 & 1.0442 & 1.0890 \\
			200 & 1.1260 & 1.2418 & 1.3552 & 1.4663 \\
			500 & 1.3233 & 1.5950 & 1.8487 & 2.0927 \\
			1000 & 1.4701 & 1.8800 & 2.2689 & 2.6477 \\
			\bottomrule[1.5pt]
		\end{tabular}\label{tab2}
	\end{table}	
	
	When there is no electric holding field, $V(z)=-\Lambda e^2/z$ is the image potential. The decay rates can be calculated from the energy spectrum (\ref{eq1}) and the wave function (\ref{eq2}). As shown in the first row of Table~\ref{tab2}, the decay rate decreases with increasing $n$. This is because highly-excited states are further away from the surface. Since ripplons represent the height variations of the helium surface, the couplings between the highly-excited electrons and ripplons are weak.
		
	In experiments, there is usually an electric holding field $E_z$ applied perpendicular to the liquid surface. By considering $E_z$ as an uniform field, the corresponding potential can be expressed as $eE_z z$. As shown in the lower four rows of Table~\ref{tab2}, the decay rates of the excited states under different electric holding fields are calculated from the eigen values and wave functions, which are solved numerically. The energy level spacings between higher-laying Rydberg states enlarge with increasing $E_z$. Meanwhile, the SE in the holding field is closer to the liquid surface than that without the holding field, especially for the highly-excited states. Thus, for the experimental configuration with a considerable electric holding field, the decay rates are significantly different from the zero-field case according to Eq.~(\ref{eqA2}). 
	
	\section{Eigenvalues from the perturbation method}
	\label{sec:AppB}
	
	Assuming that $E_n=-x/2$, the secular equation of Eq.~(\ref{eq6}) is
	\begin{align}
		x[x^2+2(\delta_1+\delta_2)x+4\delta_1\delta_2-\Omega^2]-2\Omega_1^2\delta_2=0,
	\end{align}
	where $\Omega=\sqrt{\Omega_1^2+\Omega_2^2}$. To solve the secular equation, we assume $\delta_2$ as a perturbation term and
	\begin{align}
		x_j\approx x_{0j}+A_j\delta_2.
	\end{align}
	
	The zero-order terms satisfies
	\begin{align}
		x_{0j}[x_{0j}^2+2(\delta_1+\delta_2)x_{0j}+4\delta_1\delta_2-\Omega^2]=0,
	\end{align}
	and the solutions are
	\begin{align}
		\begin{split}
			x_{01}&=0, \\
			x_{02}&=-\delta_1-\delta_2-\sqrt{(\delta_1-\delta_2)^2+\Omega^2}\\
			&\simeq-\delta_1-\delta_2-\Omega-\frac{\delta_1-\delta_2}{2\Omega}(\delta_1-\delta_2)\\
			&\simeq-\delta_1-\delta_2-\Omega, \\
			x_{03}&=-\delta_1-\delta_2+\sqrt{(\delta_1-\delta_2)^2+\Omega^2}\\
			&\simeq-\delta_1-\delta_2+\Omega+\frac{\delta_1-\delta_2}{2\Omega}(\delta_1-\delta_2)\\
			&\simeq-\delta_1-\delta_2+\Omega.
		\end{split}
	\end{align}
	The approximation is valid when $\delta_1, \delta_2 \ll \Omega $. Thus, the secular equation is transformed into
	\begin{align}
		(x-x_{01})(x-x_{02})(x-x_{03})-2\Omega_1^2\delta_2=0.
	\end{align}
	Inserting $x_j\approx x_{0j}+A_j\delta_2$, we can obtain
	\begin{align}
		\begin{split}
			A_1&=\frac{2\Omega_1^2}{(x_{01}-x_{02})(x_{01}-x_{03})}\simeq-\frac{2\Omega_1^2}{\Omega^2}, \\
			A_2&=\frac{2\Omega_1^2}{(x_{02}-x_{01})(x_{02}-x_{03})}\simeq\frac{\Omega_1^2}{\Omega^2}, \\
			A_3&=\frac{2\Omega_1^2}{(x_{03}-x_{01})(x_{03}-x_{02})}\simeq\frac{\Omega_1^2}{\Omega^2},
		\end{split}
	\end{align}
	and the eigenvalues to the first-order approximation as
	\begin{align}
		\begin{split}
			E_1&\simeq\frac{\Omega_1^2}{\Omega^2}\delta_2, \\
			E_2&\simeq\frac{\Omega}{2}+\frac{\delta_1}{2}+\frac{\Omega_2^2}{2\Omega^2}\delta_2, \\
			E_3&\simeq-\frac{\Omega}{2}+\frac{\delta_1}{2}+\frac{\Omega_2^2}{2\Omega^2}\delta_2.
		\end{split}
	\end{align}

	
	%

\end{document}